# Numerical Accuracy Comparison of Two Boundary Conditions Commonly used to Approximate Shear Stress Distributions in Tissue Engineering Scaffolds Cultured under Flow Perfusion

Olufemi E. Kadri,[1] Cortes Williams III,[2] Vassilios Sikavitsas,[2] and Roman S. Voronov[1,†]

[1] Otto H. York Department of Chemical, Biological and Pharmaceutical Engineering, New Jersey Institute of Technology, Newark, NJ 07102, USA

[2] Stephenson School of Biomedical Engineering, The University of Oklahoma Norman, OK 73019, USA

## Abstract

Flow-induced shear stresses have been found to be a stimulatory factor in pre-osteoblastic cells seeded in 3D porous scaffolds and cultured under continuous flow perfusion. However, due to the complex internal structure of the scaffolds, whole scaffold calculations of the local shear forces are computationally-intensive. Instead, representative volume elements (RVEs), which are obtained by extracting smaller portions of the scaffold, are commonly used in literature without a numerical accuracy standard. Hence, the goal of this study is to examine how closely the whole scaffold simulations are approximated by the two types of boundary conditions used to enable the RVEs: "wall boundary condition" (WBC) and "periodic boundary condition" (PBC). To that end, Lattice-Boltzmann Method fluid dynamics simulations were used to model the surface shear stresses in 3D scaffold reconstructions, obtained from high resolution microcomputed tomography images. It was found that despite the RVEs being sufficiently larger than 6 times the scaffold pore size (which is the only accuracy guideline found in literature), the stresses were still significantly under-predicted by both types of boundary conditions: between 20 and 80% average error, depending on the scaffold's porosity. Moreover, it was found that the error grew with higher porosity. This is likely due to the small pores dominating the flow field, and thereby negating the effects of the unrealistic boundary conditions, when the scaffold porosity is small. Finally, it was found that the PBC was always more accurate and computationally efficient than the WBC. Therefore, it is the recommended type of RVE. Overall, this work provides a previously-unavailable guidance to researchers regarding the best choice of boundary conditions for RVE simulations. Furthermore, it lays down the foundation for recovering more accurate scaffold stress estimates from the RVE approximations.

**KEYWORDS:** Computational fluid dynamics, surface shear stresses; Lattice Boltzmann Method; tissue engineering, wall boundary condition, periodic boundary condition, representative volume element

# 1. Introduction

Incidences of bone disorders constitute a significant economic burden to societies globally. In the United States alone, over $213 billion is the total annual cost (direct and indirect) of treating the estimated 126.6 million people affected by musculoskeletal disorders.[1] Unfortunately, with an increasingly obese and ageing population, this trend is expected to continue further. Current approaches for replacing the damaged bone tissues include the use of bone grafts (i.e., autografts or allografts). However, these methods have several shortcomings, limited availability and risk of disease transmission.[2-4] To address those disadvantages, bone tissue engineering has emerged as an alternative regenerative strategy.

In bone tissue engineering, a combination of osteo-inductive biological factors, mesenchymal stem cells obtained from patients' own bone marrow and porous biodegradable scaffolds are used. Typically, the process involves seeding the cells within the 3-D scaffolds, followed by culturing under flow in perfusion bioreactors. The flow is a necessary part of the culture, because the stimulatory shear that it imposes on the stem cells mimics the natural microenvironment in bone canaliculi.[5, 6] Moreover, it has been shown to promote tissue regeneration.[7-10] Thus, the applied shear stresses should be within the physiological range required for stimulation: 0.1 - 25 dynes/cm$^2$,[11-13] because excessive shear of 26-54 dynes/cm$^2$ can cause cell lysing and/or detachment from the scaffold.[14, 15]

Therefore, the ability to predict the shear stress distribution in different scaffold micro-architectures can provide insight into whether or not a particular scaffold design will promote tissue growth. Moreover, when used in conjunction with the latest advances in 3D microfabrication technologies, such predictive capabilities can be used to create optimized scaffold geometries. Unfortunately, however, the complex internal structure of the porous scaffolds makes estimation of the required shear stresses via experimental or analytical techniques impractical. Hence, computational fluid dynamics models, based on either idealized pore geometries [6, 8, 16-20] or actual scaffold images,[12, 21-33] are commonly utilized.

The latter is the more realistic approach since it is based on the actual microscopic pore structures, which are typically obtained via a 3D scanning technique such as micro-computed tomography (μCT). Yet, due to the computationally intensive nature of the scaffold reconstructions resulting from such high-resolution imaging, researchers are forced to resort to implementing approximations.[12, 21-33] For example, rectangular "representative volume elements" (RVE) are cut from whole scaffolds and implemented in conjunction with various boundary conditions along the artificially created periphery. Two common types of boundary conditions that are typically implemented for this purpose are the "wall boundary condition" (WBC) [8, 12, 19, 21, 22, 26, 27, 29-33] and the "periodic boundary condition" (PBC).[18, 20, 23-25, 28] In the former case, the RVE is surrounded by solid walls in the non-flow directions, while the latter is an application of periodicity in all three dimensions.

Although these approaches save on computation time, it is not obvious how accurate the resulting shear stresses are, or which of the two boundary conditions yields the better results. Consequently, the RVE approach is commonly questioned by journal reviewers, as no standards or guidance regarding their use exist. We have found only one publication that investigated the accuracy of the RVE-WBC, as compared to the whole scaffold simulation.[21] Here, a guideline was provided stating that for scaffolds with a homogenous pore distribution *the domain size should be at least 6 times the average pore size*. However, this suggestion was made based on *average* wall stresses only, while in reality the *spatial distribution* of the stresses is also important to tissue growth. For example, the cells within the scaffold migrate around in a nonrandom manner, are therefore more likely to experience stresses at some preferred locations. Furthermore, only scaffolds prepared using the same fabrication technique were studied, though two different materials were used in their manufacturing. Nonetheless, the scaffold's structure depends more on the fabrication method than it does on the material.[23, 24, 34] Moreover, just a single scaffold sample was used for each type of the material. Hence, a more thorough investigation of the RVEs' accuracy is warranted, especially given that no PBC studies were found at all.

Therefore, in this work we set out to quantify how the two relevant boundary conditions compare against each other, when applied to scaffolds manufactured using different fabrication methods, and for a large number of samples with varying porosities. To achieve this, an in-house Lattice-Boltzmann Method (LBM) code is used to simulate fluid flow through reconstructions of salt-leached foam and nonwoven fiber mesh scaffolds, all of which are imaged using µCT. The overall computational approach used in this work is summarized in **Figure 1**. Ultimately, the accuracy of the spatial stress distributions is reported for both the RVE-WBC and the RVE-PBC. Finally, a descriptive statistical analysis is used to demonstrate the accuracy and computational efficiency differences between the two RVE approaches.

## 2. Materials and Methods

### 2.1. Scaffold Preparation and 3D Imaging

Two types of porous scaffolds, salt-leached foam and nonwoven fiber scaffolds, were produced from poly-L-lactic acid using methods described in detail elsewhere.[23] The scaffolds were then scanned via µCT using a ScanCo VivaCT40 system (ScanCo Medical, Bassersdorf, Switzerland) to obtain 10µm resolution, 2D intensity image slices (see **Figure 2**) at the optimum settings of 88 µA (intensity) and 45 kV (energy). The acquired X-ray images were filtered for noise reduction and assembled into 3D reconstructions of the scaffolds using a custom Matlab code (MathWorks Inc., Natick, MA). The scans were segmented using a thresholding technique, which resulted in the porosity of scaffolds being within 1% of experimental measurements.[23-25]

### 2.2. Simulation Domain and Boundary Conditions

For the non-RVE calculations, each simulation domain was composed of a *whole* scaffold placed inside of a pipe (see **Figure 3**-**A**). This is meant to mimic the cassette holder that typically fixes the scaffold in the perfusion bioreactors. The pipe's length was taken to be approximately 10 times greater than the scaffold thickness, in order to avoid periodicity artifacts and ensure that a uniform parabolic profile is developed before flow reaches the scaffolds. Simulations were performed for a flow rate of 0.15 mL/min. This is considered a suitable flow rate for mechanical stimulation in many perfusion bioreactors.[9, 35]

For the RVE calculations, the simulation domain was obtained by extracting the *largest* rectangle that could possibly be inscribed into the circular whole scaffold (see **Figure 2**-**LEFT**). We chose the largest domain possible, in order to examine the *best-case scenario* produced by the RVEs. Subsequently, the resulting domain sizes (see **Figure 2**-**RIGHT**) were all between 8 - 14 times the scaffolds pore sizes, which is significantly greater than the suggested minimum for RVEs.[21]

For the RVE-PBC calculations, periodicity was applied in all three directions (see **Figure 3**-**B**) to approximate an infinite domain representing the full scaffold. On the other hand, for the RVE-WBC, the scaffold was surrounded by solid walls in the non-flow directions (see **Figure 3**-**C**). In addition, an entrance length equal to half of the scaffold's thickness in the flow direction was added, in order to stay consistent with the previous analysis of this boundary condition type.[21] In both implementations, the total flow rate through the RVE was decreased proportionally to the cutout size, in order to compensate for the reduction in the cross-sectional area available to flow relative to the whole scaffold.

## 2.3. Fluid Flow Modeling: Lattice Boltzmann Method

LBM was chosen for the present application, because it is especially appropriate for modeling pore-scale flow through porous media (such as scaffolds) due to the simplicity with which it handles complicated boundaries.[23-25, 28, 36-38] A previously developed custom-written, in-house code was used in this work.[23-25, 36, 38-42] The D3Q15 lattice [43] in conjunction with the single-relaxation time Bhatnagar, Gross and Krook [44] collision term approximation was used to perform simulations. The no-slip boundary condition was applied at solid faces using the "bounce-back" technique.[45] To take advantage of the inherent LBM parallelizability, domains were decomposed using message passing interface.[36, 38] Simulation convergence was defined as when average and highest velocities computed for the simulation domain vary by less than 0.01% for two consecutive time steps. The code has been validated for several flow cases for which analytical solutions are available: forced flow in a slit, flow in a pipe and flow through an infinite array of spheres.[24, 36]

## 2.4. Surface Stress Calculations and Error Analysis

Shear stress on the surface of the scaffolds was calculated following a scheme suggested in,[28] where the full shear stress tensor is calculated first, and then the maximum eigen value is evaluated using a Jacobi iteration technique. The cell culture media was assumed a Newtonian fluid, and the shear stresses at every location within the scaffolds were estimated using:

$$\underline{\underline{\sigma}} \approx \mu \left(\frac{1}{2}\right)\left(\nabla U + \nabla U^T\right) \tag{1}$$

Where $\underline{\underline{\sigma}}$ is the shear stress tensor, and $U$ is local velocity vector. The fluid dynamic viscosity was 0.01 g/cm s, which is close to that of α-MEM supplemented with 10% FBS typically used in cell culturing experiments.[46] Velocity vectors used in calculations were derived for the specified flow rate. Computed shear stress values are the largest eigenvalues of $\underline{\underline{\sigma}}$. Stress maps generated using Tecplot 360 EX 2017 (Tecplot Inc., Bellevue, WA USA) were used to visualize localized shear stress values on the scaffolds. For the accuracy comparisons, the error at every fluid surface node was calculated as follows:

$$\varepsilon_t = 100\% * \left|\frac{True\ (Whole\ Scaffold)Stress - Approximated\ (RVE)\ Stress}{True\ (Whole\ Scaffold)\ Stress}\right| \tag{2}$$

## 3. Results

In order to compare the RVEs' performance relative to each other, as well as to that of the whole scaffold simulations, we performed image-based LBM modeling using a flow rate typically encountered during artificial tissue culturing conditions. **Figure 4** shows representative results for *whole* scaffolds of two different architectures: salt-leached foam (**Figure 4**-LEFT) and nonwoven fiber-mesh (**Figure 4**-RIGHT). These very computationally-intensive models exemplify the best stress estimates, because they are the most representative of the actual flow perfusion culturing conditions. However, they take approximately 30,000-40,000 LBM steps to converge. Unfortunately, for high-resolution scaffold images (e.g., μCT), this can translate to weeks of waiting for results, even on a large supercomputer. The RVEs on the other hand, take only a fraction of that to converge, and therefore, have a higher computational efficiency. This is shown in **Figure 5**, where both axes are normalized by the corresponding whole scaffold values. From this figure, it is apparent that the RVEs take roughly 3-6 times less LBM steps to converge, which translates into a significant waiting time reduction for the user. Hence, this makes the RVE approaches attractive, as long as some error can be tolerated.

Next, we visualized the stresses produced by the RVEs, in order to gauge the amount of error incurred from their use. The first question we wanted to answer was whether the stress approximations provided by the RVEs follow a *spatial* pattern similar to that observed in the whole scaffold simulations. To that end, **Figure 6**-A & D shows stresses in RVE-equivalent domains cut out from the *whole* scaffold simulations in **Figure 4**. These are presented in order to provide a one-to-one comparison for the actual RVE-WBC and RVE-PBC cutouts shown in **Figure 6**-B & E and **Figure 6**-C & F, respectively. From **Figure 6** it is apparent that the *spatial* stress patterns produced by the RVEs are indeed very similar to those obtained from the whole scaffold simulations. However, the former stresses are significantly lower when compared with the latter.

Consequently, we used Equation 2 to quantify the spatial *accuracy* of the results produced by the RVE approximations. This was done by calculating the amount by which they deviate from the *true* stress values at every surface voxel of the scaffolds. For the purposes of this comparison, the whole scaffold stresses are considered to be the *true* values in the Equation 2 calculations, while the stresses produced by the RVE are used as the *approximated* values in the same equation. **Figure 7** is a 3D overlay of the resulting spatial error patterns in the two scaffold types from **Figure 4**. From left to right, the top row shows the salt-leached foam RVE-PBC and WBC, respectively; while the bottom row shows the fiber-mesh RVEs, in the same order. It can be immediately observed from this figure that the bulk error is similar for both RVEs, varying roughly between 40 and 60% (for these particular samples). However, there is also a spatial trend in the error: in the case of the RVE-PBC it appears to *decrease* towards the cutout's periphery, while for the RVE-WBC the opposite is the case. This makes sense, given that the presence of the wall in the latter alters the flow field near the boundary. Therefore, there are significant differences in accuracies between the two RVEs; with the RVE-PBC yielding the more superior results.

Lastly, we wanted to check whether the type of the scaffold architecture has a significant effect on the surface stress accuracy. To that end, **Figure 8** plots the mean error in the RVE surface stress, relative to those from the whole scaffold simulations. We chose the abscissa to be porosity, since it also describes the scaffold morphology. From this figure, it is evident that the RVE-WBC always results in a greater error relative to that obtained from the corresponding RVE-PBC. Another observation is that the foam scaffolds always had a smaller error than the fiber scaffolds, though this is likely an effect of their lower porosity. Specifically, the error was found to increase proportionally to the latter, indicating that the porosity influences the simulation accuracy. The absolute stresses for the data in **Figure 8** are given in

## 4. Discussion

The ability to predict the shear stresses experienced by osteoblasts during culturing is crucial for optimizing *in vitro* bone tissue engineering experiments and scaffold designs. For example, such models could help reveal scaffold features that are key to inducing bone forming responses: increased nitric oxide, prostaglandin and osteopontin production.[5, 47, 48] However, there is a danger that the simpleton RVE models, commonly implemented by researchers to save on computation time, could potentially yield unrealistic results. Yet, the only guideline on accuracy that currently exists in literature [21] applies to the RVE-WBC only, while the RVE-PBC remains uncharacterized. Moreover, the guideline is based on *average* stress values, which is not truly representative of what the cells experience during an actual experiment, given that they are not uniformly distributed in the scaffolds. Hence, we explored the *spatial* variations in the fluid-induced shear stresses obtained via *both* the RVE-PBC and the RVE-WBC scaffold approximations. These were compared by benchmarking against the computationally-intensive *whole* scaffold LBM simulations, which are considered to yield the *true* stress values. Moreover, we made sure that our cutout sizes were well beyond the "6 times the average pore size" guideline suggested by,[21] in order to examine the *best-case scenario* produced by the RVEs.

Consequently, the RVE simulations were found to have a significantly lower computational overhead (see **Figure 5**), which ultimately translates to shorter waiting times for obtaining results. Furthermore, between the two boundary condition types commonly employed in the RVE simulations, the PBC was found to converge faster than the WBC. This is largely attributed to: 1) the smaller simulation domain size of the former, since the latter models typically include an entrance length that is added in order to avoid entrance effects; and 2) the flow field being considerably simpler, since there is no wall contributing to its complexity.

Moreover, the RVE-PBC was also found to be more accurate when compared to the RVE-WBC (see **Figure 7** and **Figure 8**). Specifically, the RVE-PBC produced less error towards the cut-outs' edges, while the RVE-WBC caused more. This makes sense, given that the RVE-PBC is not affected by its boundaries due the periodicity applied, while the RVE-WBC is essentially a flow bounded by a duct. The latter is also less representative of an equivalent cut-out from the whole scaffold. Furthermore, we found that for both RVE types, the error goes up with the scaffold porosity (see **Figure 8**). The likely reason for this observation is that when the scaffold pores are tight, they have a dominating effect on shaping the flow field and the boundary condition effects are negligible. Conversely, when the pores become less restrictive, the effect of the boundary condition becomes more evident and contributes to the error.

However, even for the best-case scenario (i.e., when the cut-out is >> 6 times larger than the average pore size), the deviation from the true stress values was still between 20 and 80%, for the ten scaffolds that we tested. This is significant and can lead to misleading conclusions about the efficacy of the scaffold. Therefore, caution must be taken when using the RVE approximations. Overall, however, the fact that most of the spatial *stress* patterns were preserved, despite the use of the unrealistic boundary conditions, adds legitimacy to their use. Additionally, information about the variation of *error*, such as what is provided in this manuscript, could in theory help undo some of the inaccuracies associated with the RVEs. Until that becomes possible, however, the RVE-PBC was found to be the clear winner over the RVE-WBC, on both the accuracy and computational efficiency fronts. Therefore, it is the recommended boundary condition for large-scale simulations, especially when the porosity of the scaffold is high (which is typical for tissue engineering).

A limitation of this study is that it was performed on *empty* scaffolds, without any cells or tissues in them. As mentioned earlier, the cells are likely to build tissues in preferred locations within the scaffolds. Therefore, the effects of the RVEs should be quantified in greater detail in the locations favored by the cells. Finally, in addition to calculating the stimulatory stresses, the flow field produced by the RVEs is often also used to model the influences of metabolite transport on the tissue growth.[25, 34] Therefore, although they are beyond the scope of this manuscript, the effects of the RVE cutouts and their boundary conditions on the mass transport within the scaffolds will be addressed in our future investigations.

## 5. Conclusions

In this work, we investigated the numerical differences between two types of boundary conditions, the PBC and the WBC, which are commonly employed for enabling RVE approximations in bone tissue engineering scaffold simulations. We found that, in general, both of the RVE types followed the same *spatial* surface stress patterns as the whole scaffold simulations. However, they under-predicted the absolute stress values by a considerable amount: 20 - 80%. Moreover, it was found that the error grew with higher porosity of the scaffold but did not depend significantly on its manufacturing method. Lastly, we found that the PBC always resulted in a better prediction (i.e., lower error) than the WBC. It was also more computationally efficient, due to a smaller simulation domain size requirement. Therefore, the PBC is recommended as the boundary condition of choice for the RVE approximations. Overall, these findings fill an important knowledge gap in literature regarding the accuracy of the widely used RVE approximations. Therefore, it is our expectation that they will be used to help researchers decide whether the use of the RVE approximations is justified for their application.

## 6. Acknowledgments

Financial support from Gustavus and Louise Pfeiffer Research Foundation is gratefully acknowledged. We also acknowledge the support provided by the University of Oklahoma Supercomputing Center for Education & Research (OSCER) and Texas Advanced Computing Center (TACC) at The University of Texas at Austin under allocations for granting us access to their High-Performance Computing facilities. Both have contributed to the research results reported within this paper. URLs: http://www.ou.edu/oscer.html and http://www.tacc.utexas.edu, respectively. This work used the Extreme Science and Engineering Discovery Environment (XSEDE), which is supported by National Science Foundation grant number ACI-1548562.[49] Allocations: TG-BCS170001 and TG-BIO160074.

## 7. Disclosure Statement
The authors have no competing financial interests to declare.

# Tables and Table Captions

**Table 1** Summarized stress data and LBM steps to convergence for all the studied scaffolds.

| Scaffold type | Porosity | Shear stress (dynes/cm²) | | Mean % Error | LBM steps |
|---|---|---|---|---|---|
| | | Average | Maximum | | |
| **Foam 1** | | | | | |
| Whole scaffold | | 0.078 | 1.695 | N/A | 37000 |
| RVE-PBC | 0.8830 | 0.057 | 0.620 | 27.799 | 9000 |
| RVE-WBC | | 0.045 | 0.547 | 42.462 | 12000 |
| **Foam 2** | | | | | |
| Whole scaffold | | 0.092 | 1.462 | N/A | 39000 |
| RVE-PBC | 0.9033 | 0.056 | 0.608 | 38.860 | 9000 |
| RVE-WBC | | 0.048 | 0.551 | 48.137 | 15000 |
| **Foam 3** | | | | | |
| Whole scaffold | | 0.076 | 2.035 | N/A | 38500 |
| RVE-PBC | 0.8526 | 0.054 | 0.815 | 29.537 | 9000 |
| RVE-WBC | | 0.038 | 0.649 | 49.389 | 15000 |
| **Foam 4** | | | | | |
| Whole scaffold | | 0.081 | 1.818 | N/A | 39000 |
| RVE | 0.8600 | 0.054 | 0.813 | 33.780 | 9000 |
| RVE-WBC | | 0.040 | 0.692 | 50.292 | 15000 |
| **Foam 5** | | | | | |
| Whole scaffold | | 0.074 | 1.958 | N/A | 39500 |
| RVE | 0.8573 | 0.052 | 0.957 | 29.342 | 12000 |
| RVE-WBC | | 0.042 | 0.813 | 43.062 | 15000 |
| **Foam 6** | | | | | |
| Whole scaffold | | 0.068 | 1.263 | N/A | 39000 |
| RVE | 0.853 | 0.052 | 0.790 | 24.059 | 9000 |
| RVE-WBC | | 0.041 | 0.565 | 40.097 | 15000 |
| **Fiber 1** | | | | | |
| Whole scaffold | | 0.122 | 3.420 | N/A | 36000 |
| RVE | 0.8790 | 0.065 | 0.780 | 46.762 | 6000 |
| RVE-WBC | | 0.058 | 0.920 | 52.340 | 13000 |
| **Fiber 2** | | | | | |
| Whole scaffold | | 0.134 | 2.318 | N/A | 34000 |
| RVE | 0.9268 | 0.059 | 0.835 | 56.404 | 5000 |
| RVE-WBC | | 0.042 | 0.576 | 68.975 | 10000 |
| **Fiber 3** | | | | | |
| Whole scaffold | | 0.146 | 3.298 | N/A | 36000 |
| RVE | 0.9388 | 0.061 | 0.766 | 58.146 | 6000 |
| RVE-WBC | | 0.045 | 0.641 | 69.195 | 12000 |
| **Fiber 4** | | | | | |
| Whole scaffold | | 0.127 | 4.345 | N/A | 42000 |
| RVE-PBC | 0.9770 | 0.058 | 0.721 | 54.117 | 9000 |
| RVE-WBC | | 0.047 | 0.382 | 62.909 | 14000 |

# FIGURE CAPTIONS

**Figure 1** The image-based modeling methodology used in this work. The scaffolds were first manufactured, and then they were scanned using high-resolution μCT. Afterwards, their architecture was reconstructed virtually in 3D, and imported into the LBM fluid flow solver. Finally, the LBM simulation results were used to compute the shear stresses on the scaffold surfaces.

**Figure 2** A 2D analogy of how an RVE simulation domain is cut out of a whole scaffold μCT image. LEFT – A slice representative of a cross-section through a typical fiber mesh scaffold. The white box marks the largest rectangular area that could possibly be inscribed into the circular scaffold; RIGHT – the resulting RVE cutout from the rectangular region inscribed into the whole scaffold. Grayscale color is the X-ray radio-density of the scaffold material as imaged via μCT.

**Figure 3** Simulation domain comparison between whole scaffold simulations and the RVEs. A – Cross-sectional view of the whole simulation domain (right half is omitted for clarity). The wall and the entrance length are shown in red. In all panes, the blue arrow is the flow direction; and the gray scale color is the X-ray radio-density of the scaffold material, as imaged via μCT. B - A representative RVE-PBC simulation domain. White arrows show the directions in which periodicity is applied. Note that periodicity is also applied in both x directions as well, but the arrows are omitted for clarity. C – A cross-sectional view of RVE-WBC simulation domain (right half is omitted for clarity). The wall and the entrance length are shown in red.

**Figure 4** 3D reconstructions of the two scaffold types used in this study, with the surface stress maps (color) calculated from LBM overlaid on the *μCT* images (gray scale). LEFT - Porous foam scaffold. RIGHT – Fiber Mesh scaffold. The pipe surrounding the scaffold, and stress values below 0.1dynes/cm$^2$ are omitted for clarity. Scaffold sizes are 6.14mm x 5.1mm x 5.1mm.

**Figure 5** Total number of LBM steps required to reach convergence versus scaffold volume for RVE-PBC and RVE-WBC. The vertical and horizontal axes are normalized with respect to the whole scaffold values, respectively.

**Figure 6** Surface *stress* maps overlays for the RVE-equivalents cutout from whole scaffold simulations and the two types of boundary conditions used for the RVEs. TOP ROW: Salt-leached Foam Scaffold. BOTTOM ROW: Nonwoven Fiber Mesh. LEFT COLUMN: RVE-equivalent volumes cutout from a *whole* scaffold models: CENTER COLUMN: RVE-PBCs. RIGHT COLUMN: RVE-WBCs. Sizes are 1.61mm x 3.91mm x 2.81mm and 1.10mm x 3.79mm x 2.99mm for salt-leached and nonwoven fiber scaffolds respectively.

**Figure 7** Surface *error* maps overlays for the two RVE types, relative to the whole scaffold simulation RVE equivalents. TOP ROW: Salt-leached Foam Scaffold. BOTTOM ROW: Nonwoven Fiber Mesh. LEFT COLUMN: RVE-PBC. RIGHT COLUMN: RVE-WBC. The wall is omitted from the WBC reconstructions for clarity. Top right quarter of each scaffold is removed in order to provide a view inside. Sizes are 1.61mm x 3.91mm x 2.81mm and 1.10mm x 3.79mm x 2.99mm for salt-leached and nonwoven fiber scaffolds respectively.

**Figure 8** Average error in fluid-induced surface stress resulting from the RVE calculations plotted versus scaffold porosity and boundary condition type. Orange color – WBC. Blue Color – PBC. Triangles – Salt Leached Foam. Circles – Nonwoven Fiber Mesh

# FIGURES

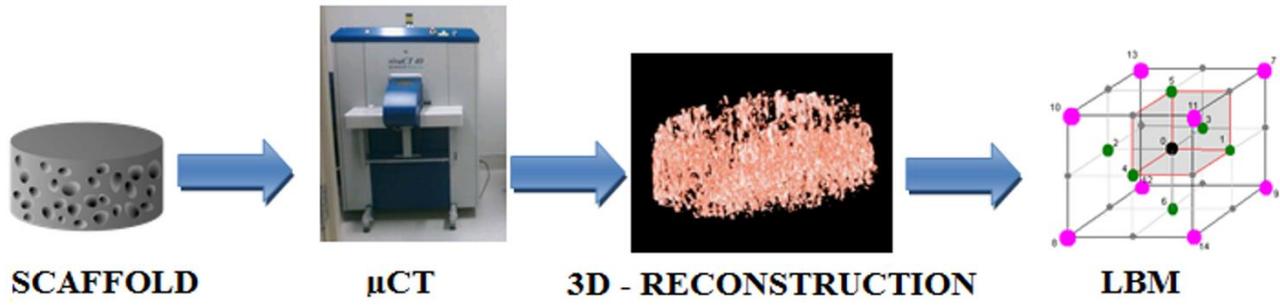

**Figure 1  Kadri et al.**

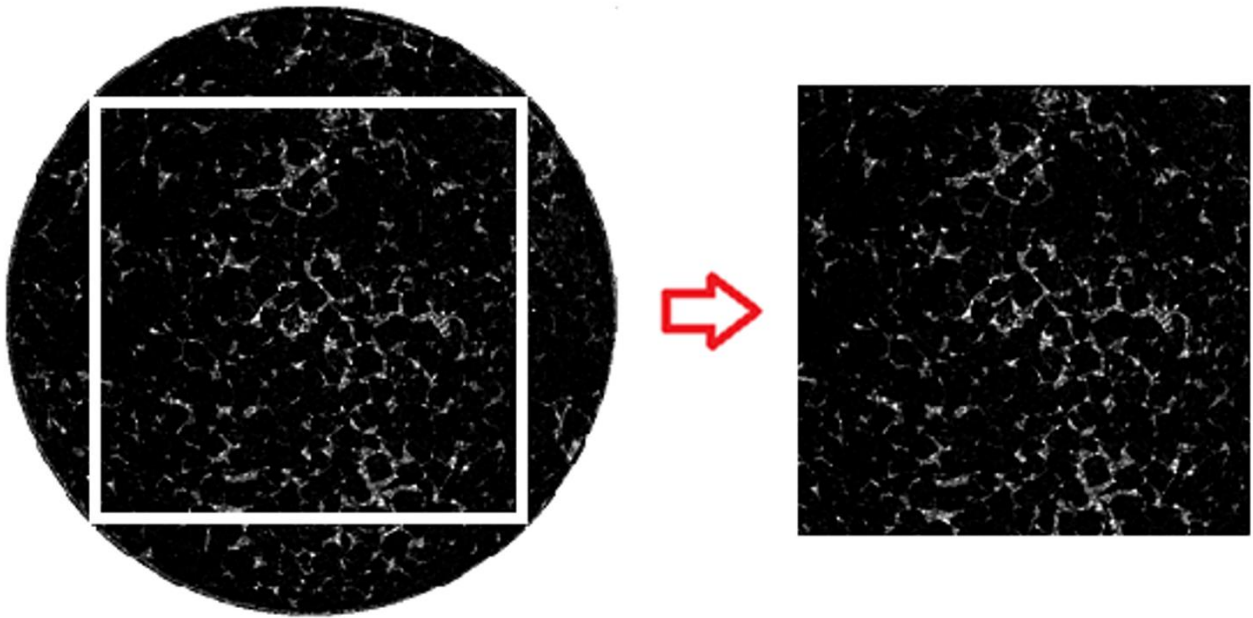

**Figure 2 Kadri et al.**

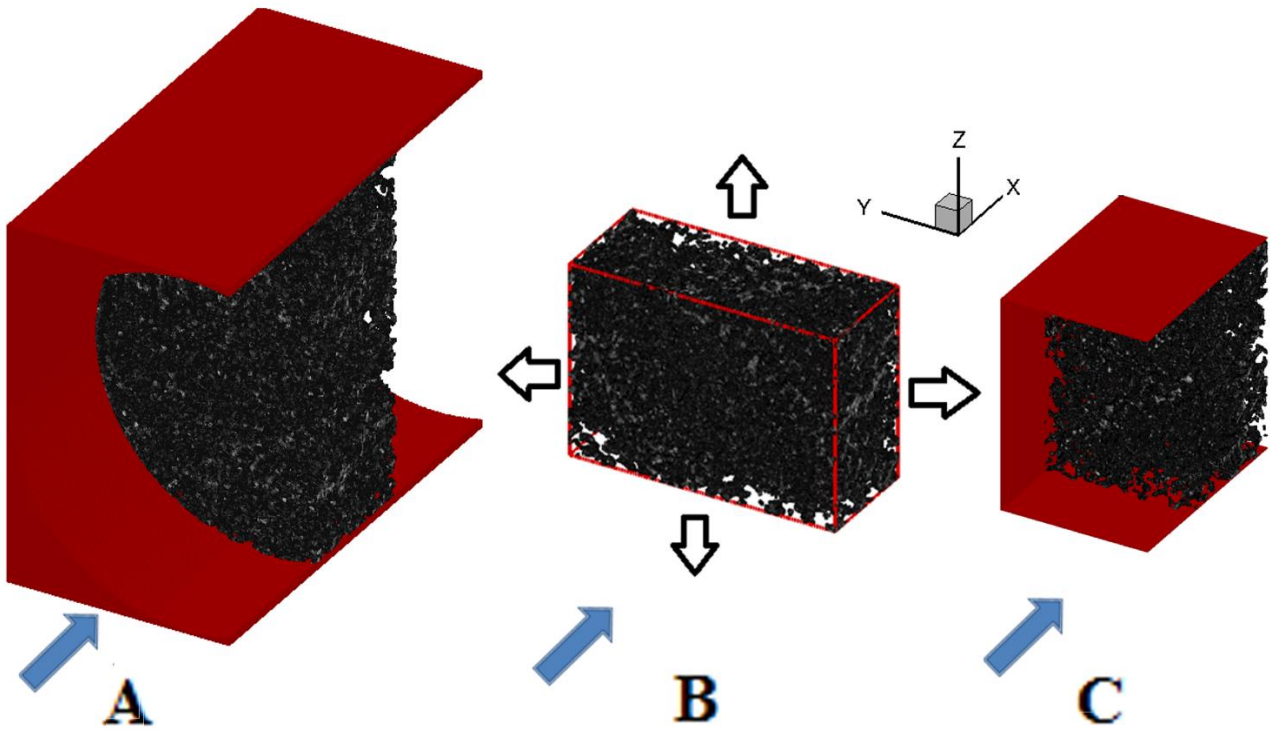

**Figure 3 Kadri et al.**

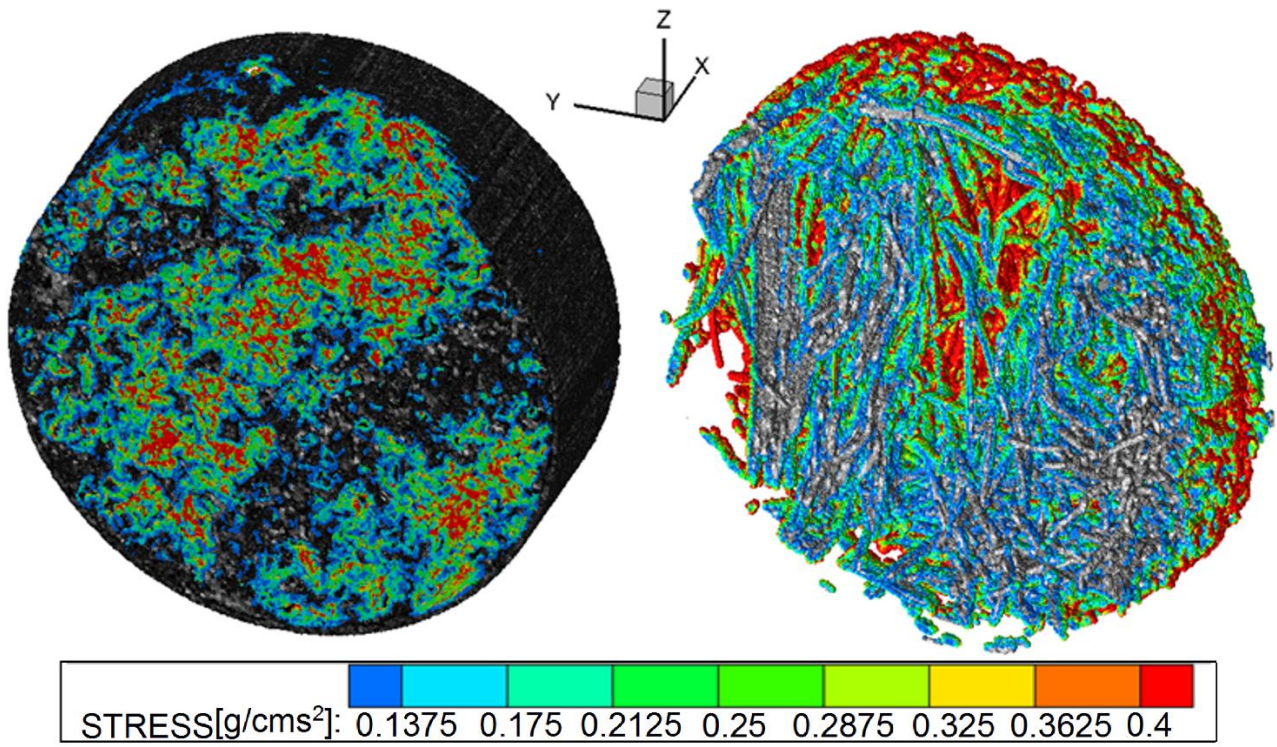

**Figure 4  Kadri et al.**

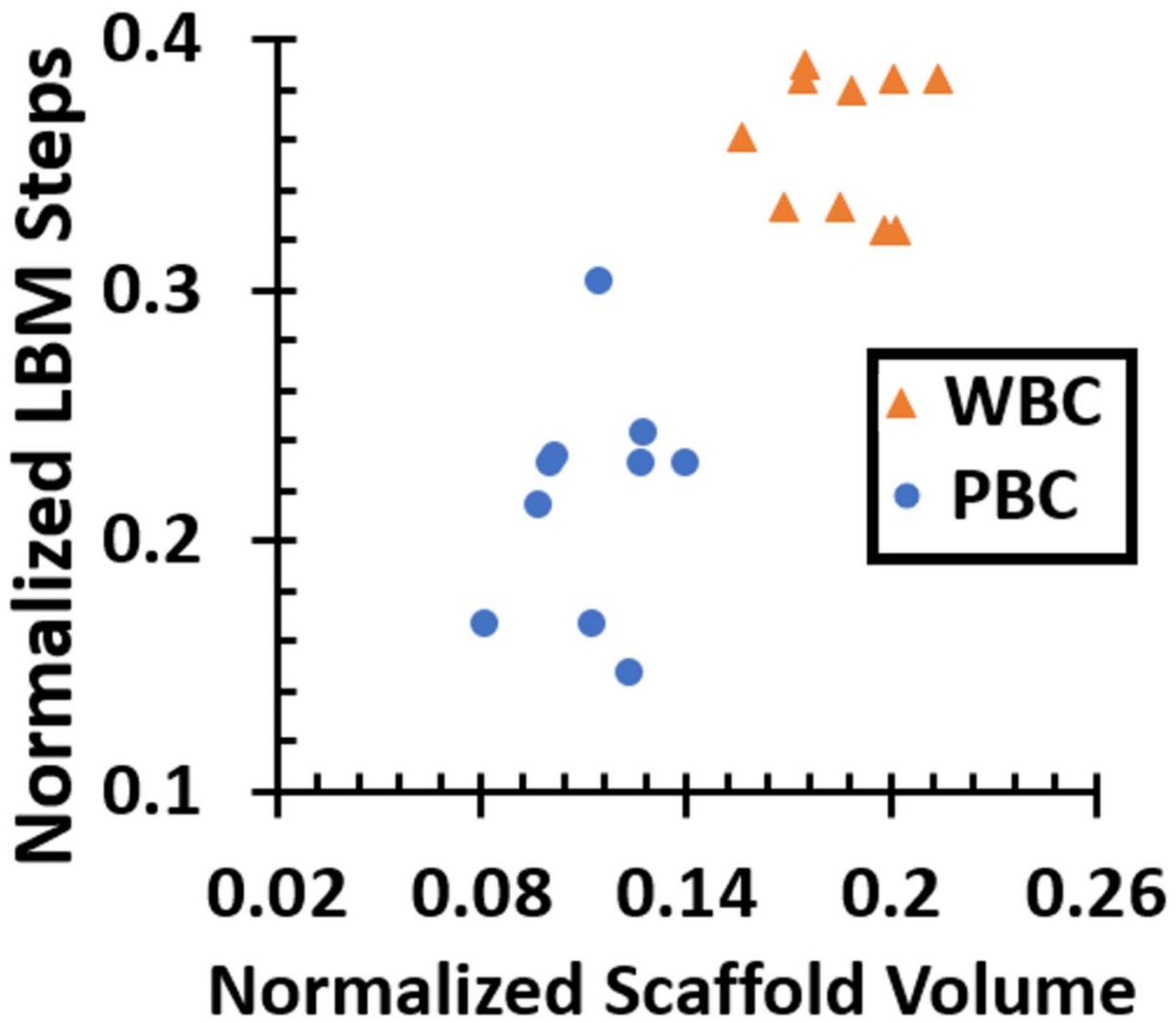

**Figure 5 Kadri et al.**

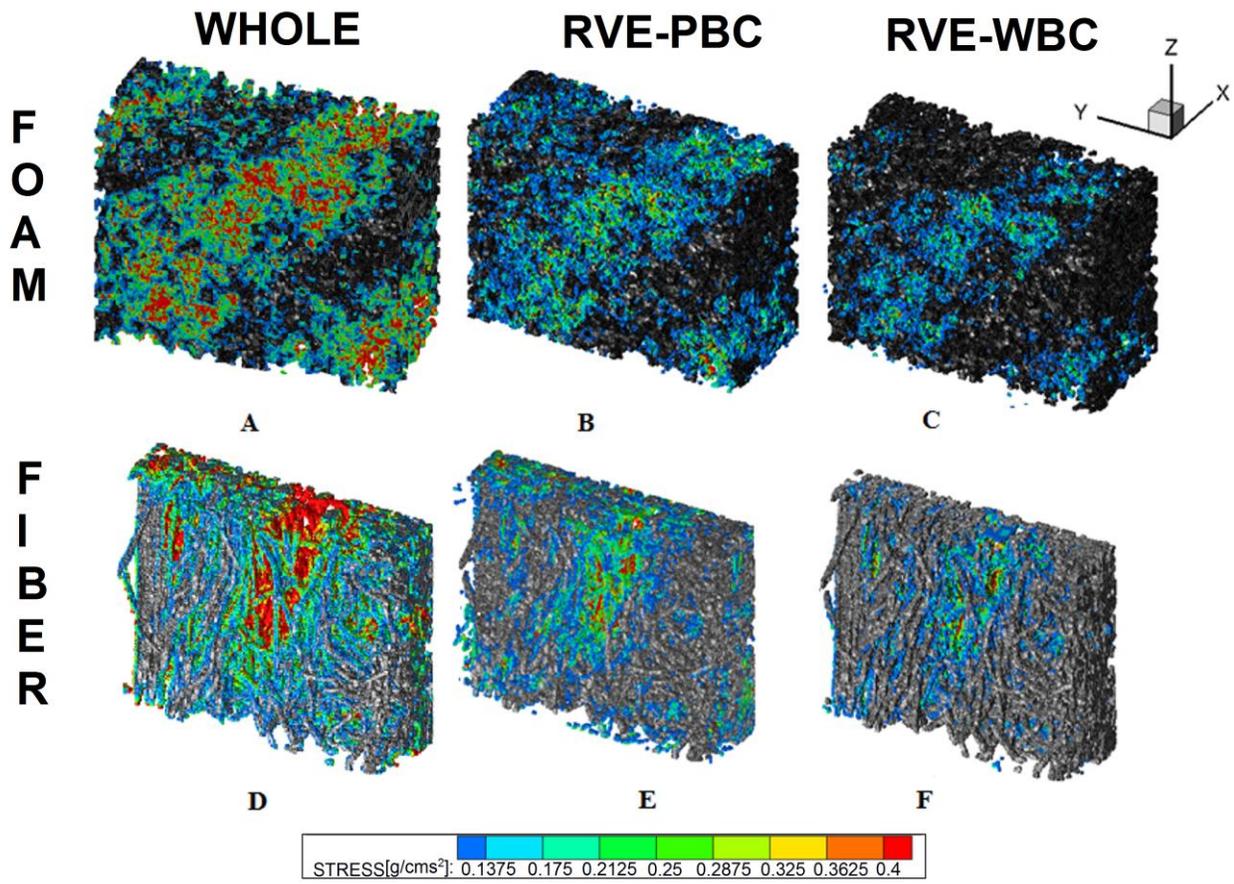

**Figure 6 Kadri et al.**

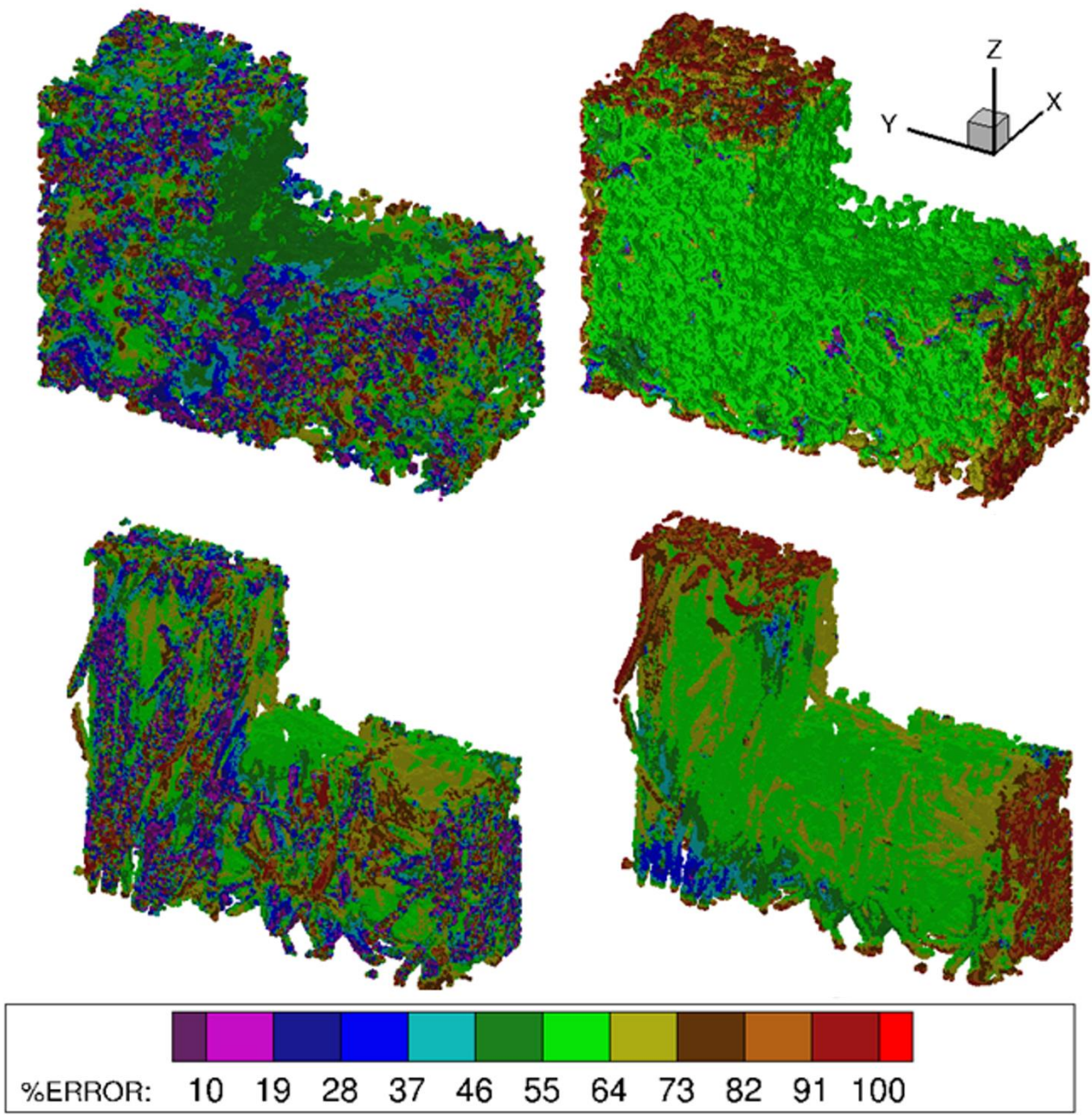

**Figure 7 Kadri et al.**

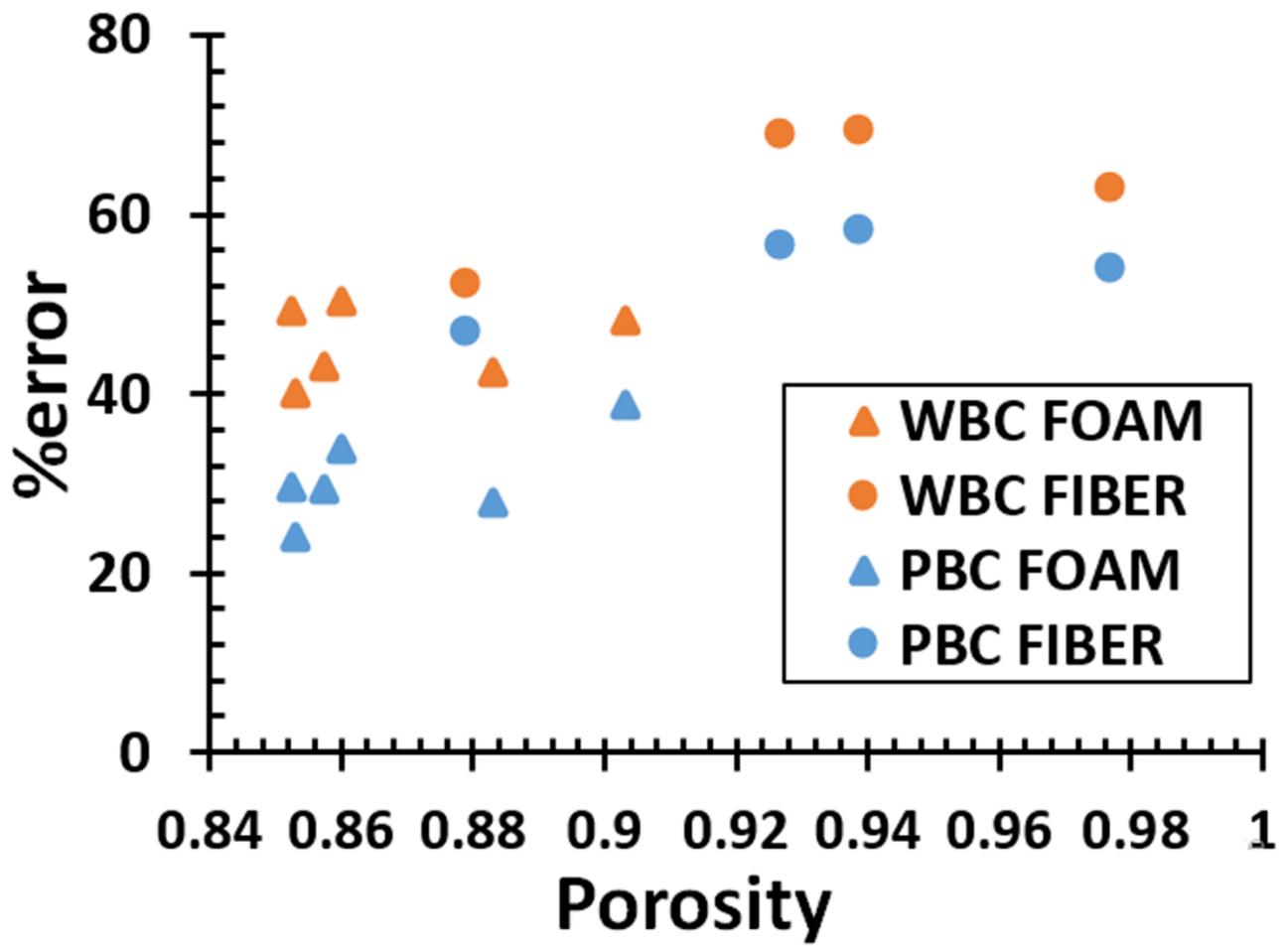

**Figure 8 Kadri et al.**